\documentclass[journal]{IEEEtran}
\ifCLASSINFOpdf
\else
\fi

\usepackage{amssymb}
\usepackage{adjustbox}
\usepackage{caption}
\usepackage{subcaption}
\usepackage{glossaries, soul}
\usepackage{pgfplots}
\usepackage{enumitem}
\usepackage{adjustbox}
\newacronym{ai}{AI}{all intra}
\newacronym{hevc}{HEVC}{high efficiency video coding}
\newacronym{jvet}{JVET}{Joint Video Experts Team}
\newacronym{cfp}{CfP}{call for proposal}
\newacronym{ipm}{IPM}{Intra prediction mode}
\newacronym{waip}{WAIP}{wide-angle Intra prediction}
\newacronym{mrl}{MRL}{multiple reference lines}

\newacronym{mip}{MIP}{matrix-based Intra prediction}
\newacronym{isp}{ISP}{Intra sub-partitions}
\newacronym{pdpc}{PDPC}{position-dependent prediction combination}
\newacronym{cclm}{CCLM}{cross-component linear model}
\newacronym{mpm}{MPM}{most probable mode}
\newacronym{ott}{OTT}{over-the-top}
\newacronym{fpga}{FPGA}{field-programmable gate array}
\newacronym{jem}{JEM}{joint exploration model}
\newacronym{mts}{MTS}{multiple transform selection}
\newacronym{dct}{DCT}{discrete cosine transform}
\newacronym{dst}{DST}{discrete sine transform}
\newacronym{ctc}{CTC}{common test conditions}
\newacronym{fft}{FFT}{fast Fourier transform}
\newacronym{bd-br}{BD-rate}{Bj\o ntegaard delta bit rate}
\newacronym{ctu}{CTU}{coding tree unit}
\newacronym{ctb}{CTB}{coding tree block}

\newacronym{rdo}{RDO}{rate-distortion optimization}
\newacronym{rd}{RD}{rate-distortion}

\newacronym{lfnst}{LFNST}{low frequency non-separable transform}

\newacronym{dpb}{DPB}{decoded picture buffer}

\newacronym{mv}{MV}{motion vector}

\newacronym{hmvp}{HMVP}{history-based motion vector prediction}
\newacronym{sbtmvp}{SBTMVP}{sub-block temporal motion prediction}
\newacronym{amvp}{AMVP}{advanced motion vector prediction}
\newacronym{amvr}{AMVR}{adaptive motion vector resolution}
\newacronym{mmvd}{MMVD}{merge with motion vector differential}
\newacronym{dmvr}{DMVR}{decoder-side MV refinement}
\newacronym{smvd}{SMVD}{symmetrical motion vector difference}
\newacronym{mvd}{MVD}{motion vector difference}

\newacronym{ip}{IP}{intellectual property}
\newacronym{rpr}{RPR}{reference picture resampling}
\newacronym{bcw}{BCW}{Bi-prediction with CU-level weights}
\newacronym{gpm}{GPM}{geometric partitioning mode}
\newacronym{ciip}{CIIP}{combined inter-intra prediction}
\newacronym{bdof}{BDOF}{bi-directional optical flow}
\newacronym{prof}{PROF}{prediction refinement with optical flow}

\newacronym{cbr}{CBR}{constant bitrate}

\newacronym{sei}{SEI}{supplemental enhancement information}

\newacronym{aps}{APS}{adaptation parameter set}
\newacronym{gci}{GCI}{general constrained information}
\newacronym{cdn}{CDN}{content delivery network}
\newacronym{sps}{SPS}{sequence parameter set}
\newacronym{vtm}{VTM}{VVC test model}
\newacronym{mtt}{MTT}{multi-type tree}
\newacronym{dft}{DFT}{discrete Fourier transform}
\newacronym{tt}{TT}{ternary-tree}
\newacronym{bt}{BT}{binary-tree}
\newacronym{qt}{QT}{quad-tree}
\newacronym{cu}{CU}{coding unit}
\newacronym{ldb}{LDB}{low delay B}
\newacronym{ra}{RA}{random access}
\newacronym{snopt}{SNOPT}{sparse nonlinear optimizer}
\newacronym{vvc}{VVC}{versatile video coding}
\newacronym{hfr}{HFR}{high frame rate}
\newacronym{hdr}{HDR}{high dynamic range}
\newacronym{asic}{ASIC}{application-specific integrated circuit}
\newacronym{sqp}{SQP}{sequential quadratic programming}
\newacronym{ampl}{AMPL}{a mathematical programming language}
\newacronym{qp}{QP}{quantization parameter}

\newacronym{lmcs}{LMCS}{luma mapping with chroma scaling}
\newacronym{ccalf}{CC-ALF}{cross-component adaptive loop filtering}
\newacronym{simd}{SIMD}{single instruction multiple data}

\newacronym{dbf}{DBF}{deblocking filter}
\newacronym{sao}{SAO}{sample adaptive offset}
\newacronym{alf}{ALF}{adaptive loop filter}
\newacronym{avc}{AVC}{advanced video coding}

\newacronym{dvb}{DVB}{digital video broadcast}
\newacronym{cmavc}{CM-AVC}{commercial module for video and audio coding}
\newacronym{atsc}{ATSC}{advanced television systems committee standards}
\newacronym{av1}{AV1}{alliance for open media (AOM) video}
\newacronym{lcevc}{LC-EVC}{low complexity enhancement video coding}
\newacronym{avs3}{AVS3}{audio video coding standard}
\newacronym{3gpp}{3GPP}{3rd generation partnership project}
\newacronym{evc}{EVC}{essential video coding}
\newacronym{gop}{GOP}{group of pictures}

\usepackage{multirow}
\newcommand{\Figure}[1] {Figure~#1}
\newcommand{\Table}[1] {Table~#1}
 \newcommand{\plh}{%
  {{\mkern-1mu\times\mkern-1.5mu}}%
}
\usepackage{amsmath,amssymb,amsfonts}
\usepackage{colortbl}
\definecolor{Gray}{gray}{1}
\usepackage{amssymb}
\definecolor{intra}{rgb}{0.84, 0.92, 0.91} 
\definecolor{tran}{rgb}{0.95, 0.59, 0.6} 
\definecolor{inloop}{rgb}{0.62, 0.79, 0.93} 
\definecolor{inter1}{rgb}{1, 0.93, 0.66} 
\definecolor{inter2}{rgb}{0.64, 0.63, 0.63} 
\definecolor{inter3}{rgb}{0.31, 0.88, 0.35} 
\definecolor{inter4}{rgb}{1, 0.93, 0.8} 

\newcommand{\addcomment}[1]{{\color{black}{#1}}}

\usepackage{hyperref}
\usepackage{pgf-pie}  
\usepackage{tabu}
\usepackage{booktabs}

\begin{document}


\title{Versatile Video Coding Standard: A Review from Coding Tools to Consumers Deployment}
%
%
%

\author{Wassim~Hamidouche~\IEEEmembership{Member,~IEEE},
        Thibaud~Biatek,
        Mohsen Abdoli,
        Edouard~François,
        Fernando Pescador~\IEEEmembership{Senior Member,~IEEE},
        Miloš Radosavljević,        
        Daniel Menard
        and Mickael Raulet~\IEEEmembership{Senior Member,~IEEE}
\thanks{Wassim Hamidouche and Daniel Menard are with Univ. Rennes, INSA Rennes, CNRS, IETR - UMR 6164, Rennes, France (e-mail: \href{mailto:wassim.hamidouche@insa-rennes.fr}{wassim.hamidouche@insa-rennes.fr} and \href{mailto:daniel.menard@insa-rennes.fr}{daniel.menard@insa-rennes.fr}).}
\thanks{Thibaud Biatek, Mohsen Abdoli and Mickael Raulet are with ATEME, Rennes, France.}
\thanks{Edouard François and Miloš Radosavljević are with InterDigital, Cesson-Sévigné, France.} 
\thanks{ Fernando Pescador is with CITSEM at Universidad Politécnica de Madrid, Madrid, Spain (e-mail: \href{mailto:fernando.pescador@upm.es}{fernando.pescador@upm.es}).}}

%
%

\markboth{IEEE Consumer Electronics Magazine, 2021}%
{Shell \MakeLowercase{\textit{et al.}}: Bare Demo of IEEEtran.cls for IEEE Journals}
%



\maketitle

\begin{abstract}
The amount of video content and the number of applications based on multimedia information increase each day. The development of new video coding standards is a challenge to increase the compression rate and other important features with a reasonable increase in the computational load. \addcomment{The joint video experts team (JVET) of the ITU-T video coding experts group (VCEG) and the ISO/IEC moving picture experts group (MPEG) have worked together to develop the versatile video coding (VVC) standard, finalized in July 2020 as the international standard 23090-3 (MPEG-I Part 3)}. This paper overviews some interesting consumer electronic use cases, the compression tools described in the standard, the current available real time implementations and the first industrial trials done with this standard.
\end{abstract}

\glsresetall

%
\IEEEpeerreviewmaketitle

\section{Introduction}
\label{intro}
\IEEEPARstart{T}{he} last two decades have witnessed exciting developments in Consumer Electronics Applications. In this framework, the multimedia applications, and more specifically those in charge of video encoding, broadcasting, storage and decoding, play a key role. Video content represents today around 82\% of the global Internet traffic according to a study recently conducted by Cisco~\cite{cisco2020cisco} and video streaming represents 58\% of the Internet traffic~\cite{Arul}. All these new trends will increase the part of video traffic, storage requirement and especially its energy footprint. For instance, video streaming contributes today to 1\% of the global greenhouse gas emissions, which represent the emissions of a country like Spain~\cite{Climate}. It is expected that in 2025, CO2 emissions induced by video streaming will reach the global CO2 emissions of cars~\cite{Climate}.

The impressive consumption of multimedia contents in different consumer electronic products (mobile devices, smart TVs, video consoles, immersive and 360$^{\circ}$ video or augmented and virtual reality devices) requires more efficient video coding algorithms to reduce the bandwidth and the storage capabilities while increasing the video quality. Nowadays, the mass market products demand videos with higher resolutions  (greater than 4K) with higher quality (HDR or 10-bit resolution) and higher frame rates (100/120 frames per second). All of these features must be integrated in devices with low resources and limited batteries. Therefore, a balance between complexity of the algorithms and efficiency in the implementations is a challenge in the development of new consumer electronic devices.

Taking into account this situation, the Joint Collaborative Team on Video Coding (JCT-VC) of ITU-T and the JCT group within ISO/IEC started working in 2010~\cite{JVET} on the development of more efficient video coding standard. An example of the success of this collaboration was the \gls{hevc}~\cite{HEVC} standard. This latter~\cite{HEVC} reduces the bit-rate of the previous video standard \gls{avc}~\cite{AVC} in a 50\% for similar visual quality~\cite{7254155}. Presently, Versatile Video Coding ~\cite{VVC} is the most recent video standard and therefore the one that defines the current state-of-the-art. The challenge of this new video standard is to ensure that ubiquitous, embedded, resource-constrained systems are able to process in real-time to the requirements imposed by the increasingly complex and computationally demanding consumer electronics video applications.

\Gls{vvc}~\cite{VVC, 9503377} has important improvements compared to its predecessors, although it is also based on the conventional hybrid prediction/transform video coding design scheme. \Gls{vvc} has achieved up to 50\%~\cite{8954562, bonnineau2021perceptual}  bitrate reduction compared to \gls{hevc} by implementing a set of new tools and features distributed over the main modules of the traditional hybrid prediction/transform coding scheme. On the other hand, the complexity of both encoder and decoder has been increased~\cite{9399488} as is explained in further sections of this paper.

The \gls{vvc} standard has been published and, at present, several research institutions and companies are working on efficient implementations that will be included in new consumer electronics devices very soon. This paper reports in further sections some efficient implementations and trials done recently in real scenarios.

The remainder of this paper is organized as follows. Section~\ref{sec:usecases} describes some use cases and the integration of the standard with other standards included in the video ecosystem. Sections~\ref{sec:vvctools} and~\ref{sec:vvccomp} outline the basic tools of the \gls{vvc} standard and the complexity of the algorithm, respectively. In Section~\ref{sec:RTcodec}, some state-of-the-art \addcomment{implementations targeting consumer electronic devices} are reported and, first commercial trials are then presented in Section~\ref{sec:trials}. Finally, Section~\ref{sec:conc} concludes the paper.


\section{Use-cases and application standards integration}

\label{sec:usecases}
    
\begin{figure}
    \centering
    \includegraphics[scale=0.26]{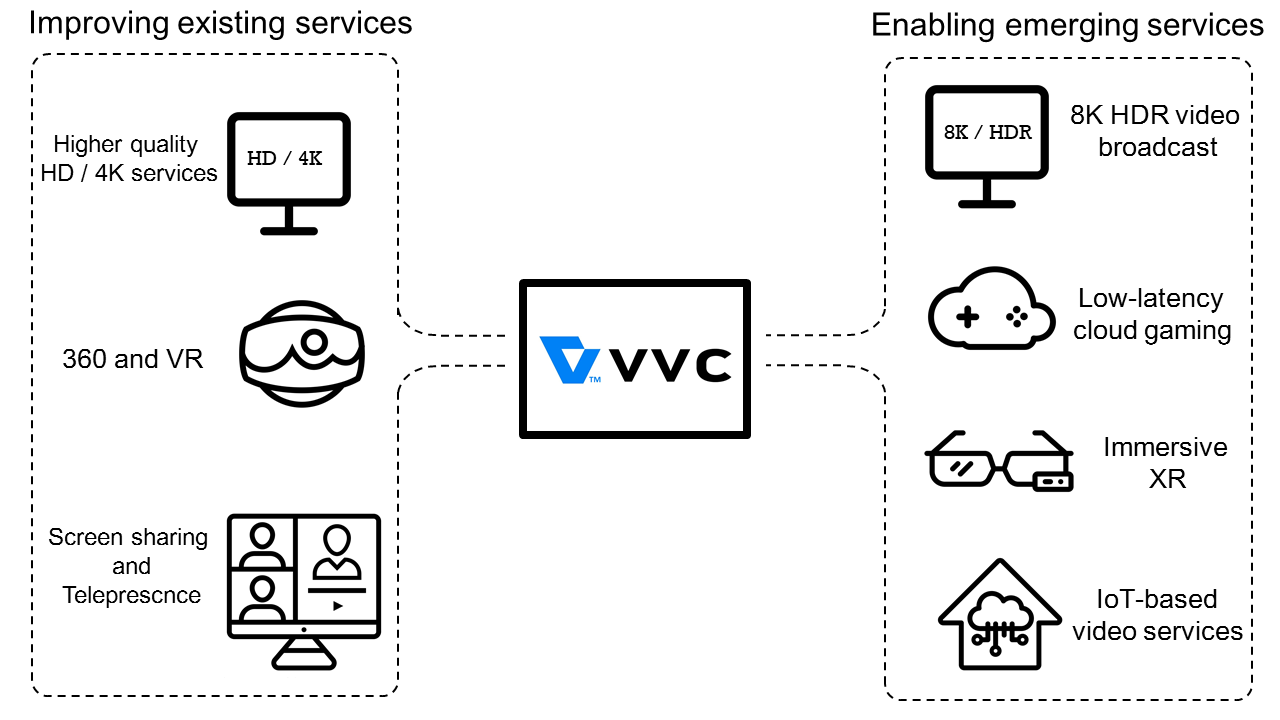}
    \caption{Potentials of \gls{vvc} from two different points of view: improving existing services and enabling emerging ones.}
    \label{fig:applications}
\end{figure}
The need for more efficient codecs has arisen from different sectors of the video delivery ecosystem, as the choice of codec plays a critical role in their success during the coming years. This includes different applications on different transport mediums and \gls{vvc} is consistently being considered as one of the main options.  The \addcomment{ \gls{vvc} standard covers a significantly} wider range of applications, compared to previous video codecs. This aspect is likely to have a positive impact on the deployment cost and interoperability issue of solutions based on \gls{vvc}. Thanks to its versatility and high capacity of addressing the upcoming compression challenges, \gls{vvc} can be used both for improving existing video communication applications and enabling new ones relying on emerging technologies, \addcomment{which are} illustrated in Figure~\ref{fig:applications}.

To properly address market needs and be deployed at scale, \gls{vvc} shall be referenced and adopted by application-oriented standards developing organization (SDO) specifications. Organizations such as \gls{dvb}, \gls{3gpp} or \gls{atsc} are defining receivers' capabilities for broadcast and broadband applications and are thus critical to foster \gls{vvc} adoption in the ecosystem. Apart from its intrinsic performance (complexity and compression), the successful adoption of a new video codec also relies on its licensing structure.

\gls{dvb}, which is a set of international open standards for digital television, is currently working to include next generation video coding solutions in the \gls{dvb} specification. In  late  2020,  before starting the standadrization activities, \gls{dvb}  organized  a workshop  on  new  video  codecs. During this workshop, five potential codecs were presented and discussed as candidates to address \gls{dvb} customers' needs: \gls{vvc}, \gls{evc}, \gls{av1}, \gls{lcevc} and \gls{avs3}. Since then, the commercial and technical work on inclusion of new video codec in the \gls{dvb} toolbox has started and is in progress, with a new TS-101-154 specification expected in 2022.

\addcomment{\gls{dvb}, which is a set of international open standards for digital television, is currently working to include next generation video coding solutions in the \gls{dvb} specification. It is expected that DVB releases a new version of the TS-101-154 specification in early 2022 including support of VVC in the video codec toolbox for application up to 8K-TV~\cite{dvbcodecs}.}

Similarly, \gls{3gpp},  which specifies mobile technologies from physical layer to application layer (e.g 4G and 5G), is also investigating the adoption of new codecs for 5G applications. Currently, three videos codecs are being characterized, namely \gls{vvc}, \gls{av1} and \gls{evc}. In TR26.955~\cite{tr26.955}, these codecs are investigated for several scenarios, such as HD-Streaming, 4K-TV, Screen-Content, Messaging, Social-Sharing and Online-Gaming.

To limit the risk of reproducing the same licensing uncertainty as \gls{hevc}, \gls{vvc} has taken a different approach. First, the media coding industry forum (MC-IF) has been created to deal with all non-technical issues related to \gls{vvc} such as licensing and commercial development. Second, the specification of \gls{sei} messages has been shifted to a dedicated specification called VSEI (Versatile SEI), published as ITU-T H.274 or ISO/IEC-23002-7. Finally, \gls{vvc} has defined in its high level syntax (HLS) a structure, named \addcomment{\gls{gci}}, enabling to switch tools off in a normative way in case licensing of specific \addcomment{\gls{ip}} would be an issue \cite{VVC}.

Finally, the integration of the \gls{vvc} with all these standards and initiatives will allow its use in different consumer electronic devices \addcomment{such as mobile phones, setup-boxes and TVs}. 

\section{\gls{vvc} Coding Tools}
\label{sec:vvctools}
\begin{figure}[t]
	\centering
	\includegraphics[width=1.0\linewidth]{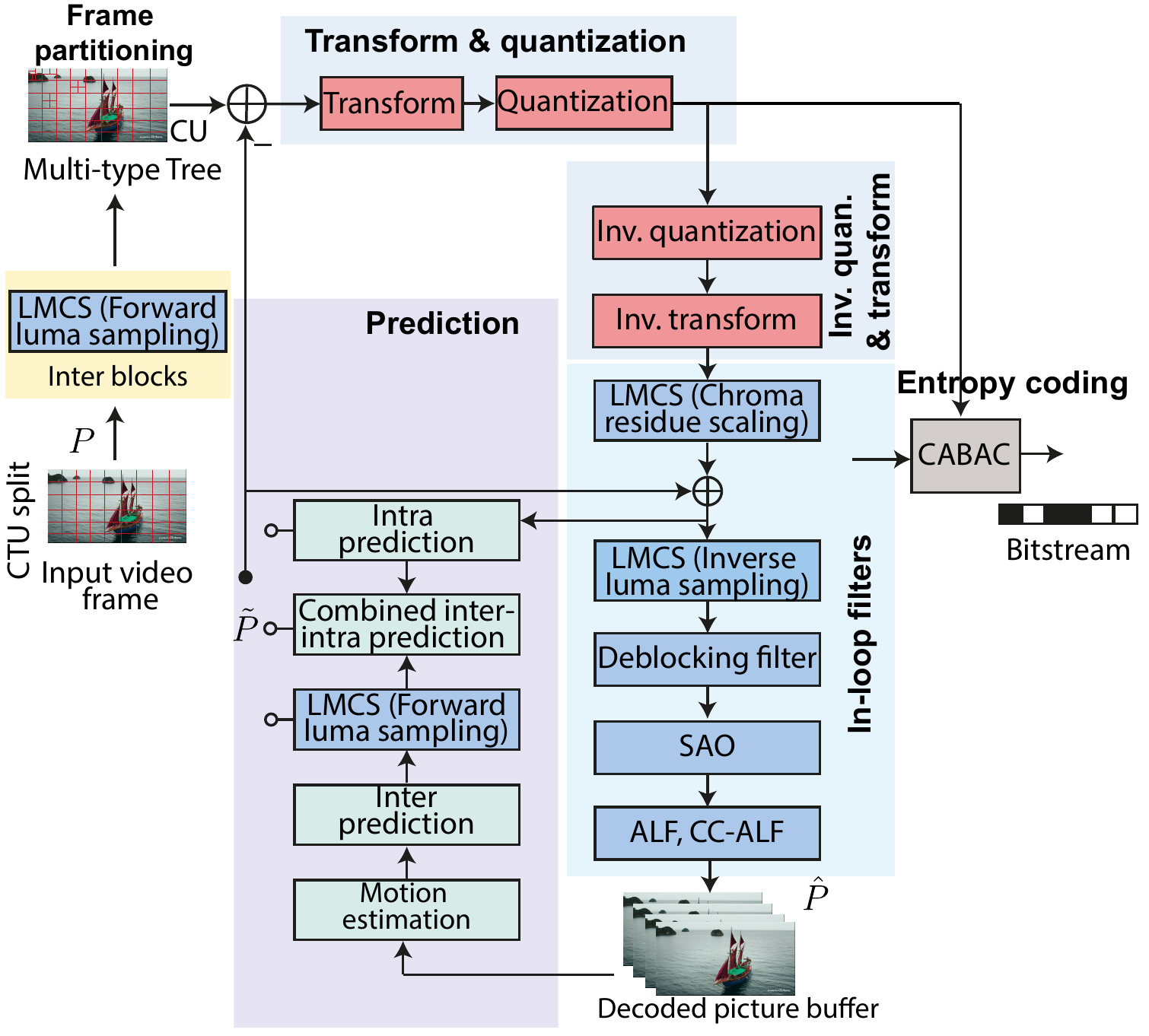}
	\caption{\gls{vvc} encoder block diagram.}
	\label{fig:VVC_encoding_diagram}
\end{figure}
In this section, we give a brief description of the \gls{vvc} coding tools to understand the improvements regarding its predecessors. Figure~\ref{fig:VVC_encoding_diagram} illustrates the block diagram of a \gls{vvc} encoder. This latter relies on a conventional hybrid prediction/transform coding. The \gls{vvc} encoder is composed of seven main blocks including 1) luma forward mapping, 2) picture partitioning, 3) prediction, 4) transform/quantization, 5) inverse transform/quantization, 6) in-loop filters, and 7) entropy coding. These seven blocks are briefly described in this section. The luma forward mapping step is normative and relies on the \gls{lmcs} tool, which is described in the in-loop filters section. \addcomment{For more exhaustive description of \gls{vvc} high level syntax~\cite{wang_high-level_2021} or \gls{vvc} coding tools and profiles, the reader may refer to the overview paper~\cite{9503377}.}             
\subsection{Picture partitioning}


The first step of the picture partitioning block splits the picture into blocks of equal size, named \gls{ctu}. The maximum \gls{ctu} size is $128\times128$ samples in \gls{vvc} and is composed of one or three \glspl{ctb} depending on whether the video signal is monochrome or contains three-color components. The \glspl{ctu} are then processed in raster scan order from top left to bottom right. In order to adapt the prediction block size to the local activity of the samples, each \gls{ctu} is then recursively split into smaller rectangular \glspl{cu}, according to the \gls{mtt} partitioning scheme. The \gls{mtt} partitioning~\cite{Huang2021partitioning} is an extension of the \gls{qt} partitioning adopted in \gls{hevc}. The blocks resulting from the partitioning process are named \glspl{cu} and may have a size between $64\times64$ and $4\times4$. In Intra slice, the luma and chroma components can be recursively split according to their own coding trees (separate luma and chroma coding trees). As in \gls{hevc}, \gls{qt} divides a \gls{cu} in four equal sub-\glspl{cu}. In addition, \gls{vvc} allows rectangular shape for \gls{cu} with its novel splits \gls{bt} and \gls{tt}. The \gls{bt} divides a \gls{cu} in two sub-\glspl{cu} while the \gls{tt} divides a \gls{cu} in three sub-\glspl{cu} with the ratio 1:2:1.  Both \gls{bt} and \gls{tt} can split a \gls{cu} horizontally or vertically. 

\subsection{Intra coding tools}
Intra coding principal takes advantage of spatial correlation existing in local image texture. To decorrelate them, a series of coding tools are provided in \gls{vvc} which are tied with partitioning and a set of \glspl{ipm} \cite{pfaff2021intra}. For the block partitioning, in addition to the principals explained in previous section, a new tool called dual-tree is introduced which allows separate partitioning trees for luma and chroma channel types. As opposed to the single-tree partitioning used in inter coding, the intra-specific dual-tree tool offers a higher level of freedom for coding decisions of chroma blocks. 


The set of \glspl{ipm} in \gls{vvc} has extended to 67 modes, compared to 35 in \gls{hevc}. This set consists of two modes of DC and planar for modeling homogeneous textures, as well as 65 directional modes for modeling angular textures. The \glspl{ipm} are coded in \gls{vvc} through an \glspl{mpm} list of six \glspl{ipm}, while a three \glspl{mpm} list is used in \gls{hevc}. For square shape blocks, these directional modes cover the same range of directional angles, with twice precision as in \gls{hevc}. Moreover, thanks to a new tool in \gls{vvc} called \gls{waip}, the set of 65 angular directional \glspl{ipm} are adaptively shifted for non-square blocks. This tool assigns additional directional modes to the longer side of non-square blocks. By doing so, prediction directions with angle greater than 45$^{\circ}$ relative to the pure horizontal or vertical modes are also possible. 


Intra prediction references are improved in \gls{vvc} to more efficiently manage noisy and less correlated neighboring samples. A reference selection tool, called \gls{mrl} has been introduced that allows the encoder to choose among three reference lines and explicitly signal the best one. Once the reference line is selected, the reference sample smoothing and the reference sample interpolation are two mechanisms in \gls{vvc} to denoise reference samples. The choice between these two mechanisms is made implicitly (\textit{i.e} without signaling) and based on block characteristics. 

\gls{pdpc} is yet another new tool in \gls{vvc} that implicitly combines the prediction signal of a block with its unfiltered and filtered boundaries. Although \gls{pdpc} is applied differently on DC, planar, horizontal/vertical and other directional modes, its general functionality can be formulated as follows:

\begin{equation} \label{eq_pdpc}
\begin{split}
 \tilde{P}(x,y) = & \lfloor  ( \omega_LR_L + \omega_TR_T \\
 & +  ( 64-\omega_L-\omega_T  ) \tilde{P}(x,y) + 32   )  / 2^6 \rfloor,
\end{split}
\end{equation}
where, $\tilde{P}(x,y)$ is the predicted sample at position of coordinates $(x,y)$, $R_T$ and $R_L$ are, respectively, reference samples aligned from the top and left to this position. Weights of these reference samples, expressed as $\omega_T$ and $\omega_L$, respectively, are determined based on the position $(x,y)$ as well as the selected intra mode. Finally, in this equation, the operator $\lfloor x  \rfloor$ is the lower integer value of a real $x$ used with a division by $2^6$ to normalize the modified prediction value into its range.

\gls{isp} tool allows splitting an intra block into two or four sub-blocks, each having its separate residual block, while sharing one single intra mode. The initial motivation behind this tool is to allow short-distance intra prediction of block with non-stationary texture, by sequential processing of sub-partitions. 
On thin non-square blocks, this scheme can result in sub-partitions coded with 1-D residual blocks, providing the closest possible reference lines for intra prediction.


The \gls{mip} in \gls{vvc} is a new tool designed by an AI-based data-driven method. \gls{mip} modes replace the functionality of conventional \glspl{ipm}, by matrix multiplication of reference lines, instead of their directional projection. The process of predicting a block with \gls{mip} consists in three main steps: 1) averaging with sub-sampling, 2) matrix-vector multiplication and 3) linear interpolation. The data-driven aspect of \gls{mip} is expressed in the second step, where a set of matrices are pre-trained and hard-coded in the \gls{vvc} specification. This set is designed to provide distinct matrices for different combinations of block size and internal \gls{mip} mode.


\Gls{cclm} is a new tool in \gls{vvc} for exploiting local correlations between luma and chroma channels. This tool is based on a similar concept in the \gls{hevc} range extension standard, where the inter-channel correlation is modeled in the residual domain. In \gls{vvc}, this modeling is carried out in the reconstructed pixel domain, in the form of: 
\begin{equation}
    \tilde{P}_c(x,y)=\alpha \, \hat{P}(x,y) + \beta,
\end{equation}
where $ \tilde{P}_c(x,y)$ is the predicted chroma value at position $(x,y)$ based on the co-located reconstructed luma value $\hat{P}(x,y)$. The model parameters $\alpha$ and $\beta$ are explicitly derived based on the relation between neighboring luma and chroma samples.

\subsection{Inter coding tools}
\begin{figure}[h]
\includegraphics[width=0.48\textwidth]{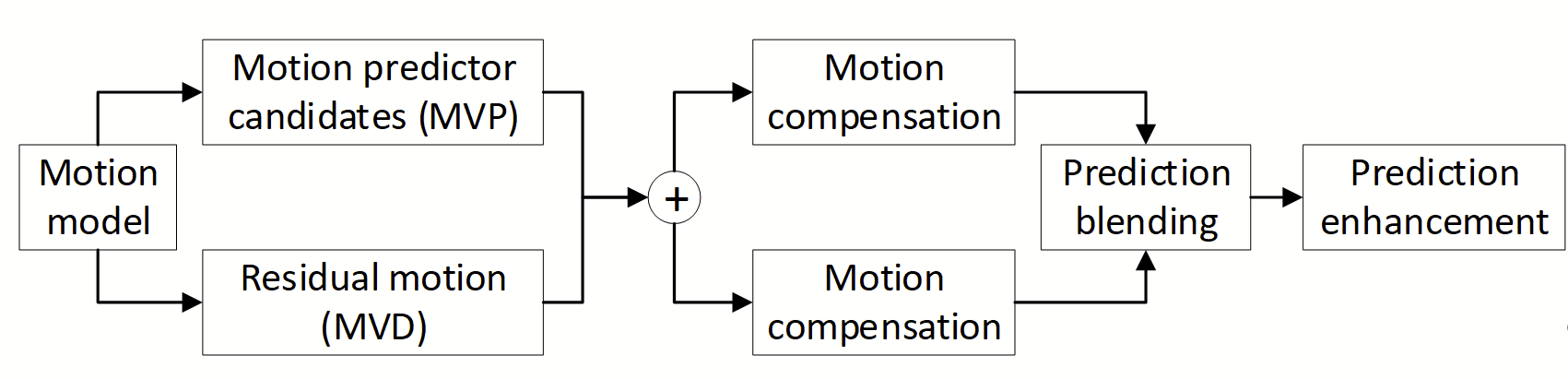}
\caption{Block diagram of the \gls{vvc} inter-prediction process.}
\label{fig:block-diagram}
\end{figure}

Inter coding relies on inter-prediction of motion and texture data, from previously reconstructed pictures stored in the \gls{dpb}. A simplified block diagram of the \gls{vvc} inter-prediction process is provided in \Figure{\ref{fig:block-diagram}}. The process first involves motion prediction, based on a list of motion data candidates. The motion prediction can be corrected by residual motion information signaled in the bitstream. The reconstructed motion vectors are then used to perform one or two motion compensations, whether the coding block is uni- or bi-predicted. When bi-prediction is performed, a blending process is then applied to mix the two motion compensated blocks. Finally, a prediction enhancement step is performed as a post-prediction filtering. The resulting predicted signal can be further corrected by a residual block signaled in the bitstream.

Two different motion models are supported in \gls{vvc}, translational model and affine model, controlled at \gls{cu}-level. The affine model can rely on 4 or 6 degrees of freedom. When a \gls{cu} is coded with affine motion, it is split into 4$\times$4 sub-blocks, whose \glspl{mv} are derived from the affine parameters of the CU. These parameters are deduced from 2 or 3 control point motion vectors positioned in the top-left, top-right and possibly bottom left corners of the \gls{cu}, depending on the number of affine parameters.  

In \gls{vvc}, the motion vectors accuracy is of 1/16th sample for luma component, and 1/32nd sample for chroma component (for 4:2:0 chroma format). An inter \gls{cu} can be coded according to three main modes: 1) skip mode, that only specifies a motion data predictor (motion vector and reference picture index) among a set of motion data candidates, and does not add any motion residual or texture residual to the predicted motion and texture block; 2) merge mode, that additionally signals a texture residual; 3) \gls{amvp} mode, that is a superset of merge mode where the whole motion information is signaled. Each of these modes uses a list of motion information candidates. In addition, the lists construction differs for the translational and affine modes. As in \gls{hevc}, the lists are built from neighboring spatial motion information of the current \gls{cu} (spatial MVs), motion information from reference pictures (TMVP). In addition, \gls{vvc} specifies two new types of motion candidates, \gls{hmvp}, and pairwise \gls{mv}. \gls{hmvp} stores in a FIFO buffer up to 5 \glspl{mv}, which allow accessing \glspl{mv} not present in the spatial or temporal neighborhood of the \gls{cu}. Pairwise \gls{mv} is an average of the two first candidates in the list. \gls{vvc} also supports a new mode, \gls{sbtmvp} that performs temporal motion field prediction of a \gls{cu} on a 4$\times$4 block granularity.
Residual motion signaling applies in \gls{amvp} mode and consists in coding a corrective motion information added to the motion information prediction. In \gls{vvc}, new residual motion coding tools are supported compared to \gls{hevc}. \Gls{amvr} allows adapting at \gls{cu}-level the coding precision of the \gls{mvd}. \Gls{smvd} applies in the case of bi-directional prediction and consists in coding a single corrective \gls{mv} applying symmetrically to \gls{mv} of each direction. In merge mode case, \gls{mmvd} mode allows slightly correcting the predicted motion information. Finally, in case of bi-directional prediction, it is also possible to refine the \glspl{mv} of a \gls{cu}, with a 4$\times$4 granularity using the \gls{dmvr} mode, that performs a motion search of limited range per 4$\times$4 sub-block.


The motion compensation is based on separable linear 8-tap filters / 16 phases for luma, and 4-tap filters / 32 phases for chroma. In \gls{vvc}, it is also possible to adaptively change the coded picture resolution, using the \gls{rpr} tool. Four different sets of interpolation filters are used depending on the motion model, block size, and scaling ratio between the current picture and the reference picture (maximum downsampling ratio 2 and upsampling ratio 8). \gls{rpr} offers new capabilities for bit-rate control to adapt to network bandwidth variations and is also applicable to scalability and adaptive resolution coding. For 360$^\circ$ video content, using equi-rectangular projection format, horizontal wrap around motion compensation allows limiting seam artifacts by performing wrapping instead of padding of the samples located at the reference picture vertical borders. 
Regarding the motion prediction blending, several new modes compared to \gls{hevc} are supported in \gls{vvc}. \Gls{bcw} is an enhanced version of the bi-prediction blending of \gls{hevc}, performing a weighted averaging of the two prediction signals $\hat{P}_0$ and $\hat{P}_1$ according to the following formula: 
\begin{equation}
\tilde{P} = \lfloor ( ( 8 – w ) \, \hat{P}_0 + w \, \hat{P}_1 + 4 ) / 2^3 \rfloor,
\end{equation}
where the weight $w$ is selected among a pre-defined set of weights. \gls{vvc} also supports \gls{gpm}, that splits a \gls{cu} into non-rectangular sub-partitions, each partition embedding a translational \gls{mv}.

\Gls{ciip} generates a mixed version of temporal prediction $\tilde{P}_{Inter}$ and spatial prediction $\tilde{P}_{Intra}$ samples according to the following formula:
\begin{equation}
\tilde{P} = ( w \,  \hat{P}_{Inter} + (1 – w ) \, \hat{P}_{Intra} ) / 4,
\end{equation}
where the mixing weight $w$ depends on the top and left CUs coding mode (intra or inter). 
The final prediction enhancement step is a new coding step in \gls{vvc} that is not present in \gls{hevc}. This step consists in slightly adjusting the prediction signal, according to two possible modes both relying on the optical flow principles. \Gls{bdof} applies in case of bi-directional prediction and consists in correcting the samples value based on the \gls{mv} and spatio-temporal gradients derived from the two reference pictures. \Gls{prof} applies to \gls{cu} coded with affine model and performs a per-sample correction that takes into account the difference between the true per-sample motion field derived from the affine model parameters, and the approximated 4$\times$4 motion field used in the actual affine motion compensation step.


\subsection{Transform and quantization} 
The transform module in \gls{vvc} is composed of two blocks namely \gls{mts} and \gls{lfnst} that perform separable and non-separable transforms, respectively~\cite{9449858}. 

{\bf \gls{mts}}: The \gls{mts} block in \gls{vvc} involves three transform types including the \gls{dct}-II, \gls{dct}-VIII and \gls{dst}-VII.  The kernels of \gls{dct}-II $C_2$, \gls{dst}-VII $S_7$ and \gls{dct}-VIII $C_8$ are derived from ~\eqref{Equ:dct-2}, \eqref{Equ:dst-7} and \eqref{Equ:dct-8}, respectively.  
\begin{equation}
C^N_{2\, i, j} = \gamma_i \sqrt{\frac{2}{N}} \cos \left (  \frac{\pi (i-1)(2j-1)}{2N}  \right ) ,
\label{Equ:dct-2}
\end{equation}
with $\gamma_i=\left\{ \begin{array}{cc}
\sqrt{\frac{1}{2}} & i=1, \\ 
1 & i\in \{2, \dots ,N \}. \end{array}
\right.$.

\begin{equation}
S^N_{7\, i,j}  = \sqrt{\frac{4}{2N+1}} \sin \left ( \frac{\pi (2i-1)j}{2N+1} \right).
\label{Equ:dst-7}
\end{equation}

\begin{equation}
C^N_{8\, i, j} = \sqrt{\frac{4}{2N+1}} \cos \left (  \frac{\pi (2i-1) (2j-1)}{2(2N+1)} \right ), 
\label{Equ:dct-8}
\end{equation} 
with $(i, j) \in \{1, 2, \dots, N \}^2$ and $N$ is the transform size. 

The \gls{mts} concept selects, for Luma blocks of size lower than 64, a set of transforms that minimizes the rate distortion cost among five transform sets and the skip configuration. However, only \gls{dct}-II is considered for chroma components and Luma blocks of size 64. The { \it sps\_mts\_enabled\_flag} flag defined at the \gls{sps} enables to activate the \gls{mts} at the encoder side. Two other flags are defined at the \gls{sps} level to signal whether implicit or explicit \gls{mts} signalling is used for Intra and Inter coded blocks, respectively. For the explicit signalling, used by default in the reference software, a syntax element signals the selected horizontal and vertical transforms.
To reduce the computational cost of large block–size transforms, the effective height $M^\prime$ and width $N^\prime$ of the coding block (CB) are reduced depending of the CB size and transform type
\begin{equation}
N^\prime =\left\{ \begin{array}{cc}
min(N, 16) & trTypeHor > 0, \\ 
min(N, 32) & \text{otherwise}. \end{array}
\right.
\label{Equ:NNzeroH}
\end{equation}

\begin{equation}
M^\prime =\left\{ \begin{array}{cc}
min(M, 16) & trTypeVer > 0, \\ 
min(M, 32) & \text{otherwise}. \end{array}
\right.
\label{Equ:NNzeroV}
\end{equation}


In~\eqref{Equ:NNzeroH} and \eqref{Equ:NNzeroV}, $M^\prime$ and $N^\prime$ are the effective width and height sizes, $trTypeHor$ and $trTypeVer$ are respectively the types of vertical and horizontal transforms (0: \gls{dct}-II, 1: \gls{dct}-VIII and 2: \gls{dst}-VII), and the $min(a,b)$ function returns the minimum between $a$ and $b$. The sample value beyond the limits of the effective $N$ and $M$ are considered to be zero, thus reducing the computational cost of the 64-size \gls{dct}-II and 32-size \gls{dct}-VIII/\gls{dst}-VII transforms. This concept is called zeroing in the \gls{vvc} specification.

{\bf \gls{lfnst}}: The \gls{lfnst} has been adopted in \gls{vvc} since the \gls{vtm} version 5. The \gls{lfnst} relies on matrix multiplication applied between the forward primary transform and the quantisation at the encoder side:
\begin{equation}
\vec{Z} = T \cdot \vec{Y},
\end{equation}
where the vector $\vec{Y}$ includes the coefficients of the residual block rearranged in a vector and the matrix $T$ contains the coefficients transform kernel. 
The inverse \gls{lfnst} is expressed in~(\ref{equ:invsep-trans}).
\begin{equation}
\vec{\tilde{Y}} = T^T \cdot \vec{\tilde{Z}}.
\label{equ:invsep-trans}
\end{equation}

Four sets of two \gls{lfnst} kernels of sizes $16\plh16$ and $64\plh64$ are applied on 16 coefficients of small blocks (min (width, height) $<$ 8 ) and 64 coefficients of larger blocks (min (width, height) $>$ 4), respectively. The \gls{vvc} specification defines four different transform sets selected depending on the Intra prediction mode and each set defines two transform kernels. The used kernel within a set is signaled in the bitstream. The transform index within a set is coded with a Truncated Rice code with rice parameter $p=0$ and $cMax =2$ (TRp) and only the first bin is context coded. The \gls{lfnst} is applied on Intra \gls{cu} for both Intra and Inter slices and concerns Luma and Chroma components. Finally, \gls{lfnst} is enabled only when \gls{dct}-II is used as a primary transform.       

\subsection{In-loop filters}
The picture partitioning and the quantization steps used in \gls{vvc} may cause coding artifacts such as block discontinuities, ringing artifacts, mosquito noise, or texture and edge smoothing. Four in-loop filters are thus defined in \gls{vvc} to alleviate these artifacts and enhance the overall coding efficiency~\cite{9399506}. The \gls{vvc} in-loop filters are \gls{dbf}, \gls{sao}, \gls{alf} and \gls{ccalf}. In addition, the \gls{lmcs} is a novel tool introduced in \gls{vvc} that performs both luma mapping to the luma prediction signal in inter mode and chroma scaling to residuals after inverse transform and inverse quantisation. The \gls{dbf} is applied on block boundaries to reduce the blocking artifacts. The \gls{sao} filter is then applied on the deblocked samples. The \gls{sao} filter first classifies the reconstructed samples into different categories. Then, for each category, an offset value retrieved by the entropy decoder is added to each sample of the category. The \gls{sao} is particularly efficient to alleviate ringing artifacts and correct the local average intensity changes. The last in-loop filters, \gls{alf} and \gls{ccalf}, perform block-based linear filtering and adaptive clipping. The \gls{alf} performs adaptive filtering to minimize the mean squared error (MSE) between original and reconstructed samples relying on Wiener filtering. A 7$\times$7 diamond filter shape is applied on luma components. The filter coefficients are derived from a 4$\times$4 block classification based on local sample gradients. According to the computed class, filter coefficients are selected from a set of filters which are whether fixed or signaled in the bitstream at the level of the \gls{aps}. Geometric transformations such as 90-degree rotation, diagonal or vertical flip may also be applied to the filter coefficients according to the block class. For chroma samples \gls{alf}, a 5$\times$5 diamond filter shape is first applied. The  chroma filter coefficients can only be signaled in the \gls{aps}. The \gls{ccalf} uses co-located Luma samples to generate a correction for chroma samples. The \gls{ccalf} is applied only on chroma samples and it is performed in parallel with the \gls{alf}.

\section{Complexity and Coding performance}
\label{sec:vvccomp}
In order to assess the benefits of the \gls{vvc} coding tools described in previous sections, a “tool off” test has been performed, consisting, for each individual tool, in evaluating the coding cost variations between an encoding setting with all tools enabled, compared to an encoding setting with all tools enabled except the tested tool. A set of 42 UHD sequences, not included in the common test sequences used during the \gls{vvc} development process, has been used. This set includes sequences of various texture complexity, motion amplitude and local variations, and frame rates (from 30 to 60 frames per second). 
The evaluation has been performed in \gls{ra} configuration, with \gls{gop} size of 32, and one intra frame inserted every 1 second, using the \gls{vvc} reference software (\gls{vtm}1\addcomment{1}.0). The evaluation focuses on the main new tools supported by \gls{vvc} and not present in \gls{hevc} (except the partitioning \addcomment{and the entropy} coding parts which are not considered in this evaluation). \addcomment{For motion coding, only tools incrementally added to the \gls{hevc} design are considered. SAO, which has the same design in HEVC and VVC, has also been evaluated}. The \gls{bd-br} metric ~\cite{ITU_Metrics} is used as estimation of the bitrate variations, using as objective quality metrics PSNR, VMAF ~\cite{VMAF} and MS-SSIM ~\cite{MSSIM}. For PSNR metric, the \gls{bd-br} variations are computed for the PSNR of each color component (Y, U, V), then a weighted \gls{bd-br} value is computed from the \gls{bd-br} of each component (using weight \addcomment{6/8} for luma, and \addcomment{1/8} for each chroma component). VMAF and MS-SSIM are only computed on the luma (Y) component. A positive value of \gls{bd-br} variation indicates a bit rate increase when disabling the tool. Encoding and decoding runtimes variations are also reported. These latest figures are provided as indicative data, but must be considered with a lot of care, since the \gls{vtm} decoder implementation is far from a real product implementation and from an efficient implementation. Detailed results are reported in~\Table{\ref{table_example}}. Tools are grouped per category ($^{\dagger}$~Intra tools, $^{\oplus}$~Transform tools, $^\otimes$~In-loop filter tools, $^\star$~\addcomment{MV coding, $^\odot$~Subblock Motion Compensation}, $^{\bullet}$~Prediction blending, $^{\diamond}$~Prediction enhancement). When all these tools are disabled, \gls{bd-br} variations are \addcomment{43.18}\% for PSNRYUV, 30.22\% for VMAF and 27.67\% for MS-SSIM, with encoding and decoding runtime \addcomment{variations} of 19\% and 49\% compared to the setting with all tools enabled (anchor). This demonstrates the substantial coding performance brought by the new \gls{vvc} tools. \addcomment{It is also observed that, for most of the evaluated tools, the reported results are very consistent with the per-tool evaluation reported by JVET document~\cite{JVET-T0013}, which demonstrates that tools performance was not optimized for the JVET test sequences. LMCS presents lower gains in this paper than in~\cite{JVET-T0013}, which tends to show that this tool requires very accurate and content-dependent tuning. Gains from SAO are small in terms of objective metrics, but this tool was considered by JVET as particular relevant for subjective quality. Also, its  implementation is extremely low-cost, which makes this tool very useful when ALF is disabled.}

\begin{table}[!t]
\renewcommand{\arraystretch}{1.3}
\caption{Per-tool \gls{vvc} performance. The \gls{bd-br} performance are reported by disabled tool in terms of PSNR, VMAF and SSIM. The relative encoding \addcomment{time} (EncT) and decoding \addcomment{time} (DecT) \addcomment{variations} are reported with respect to the anchor.}
\label{table_example}
\begin{adjustbox}{max width=0.5\textwidth}
\centering
\begin{tabular}{|l|c|c|c|c|c|}
\hline

\multirow{2}{*}{Tools}  & \multicolumn{3}{c|}{\gls{bd-br}} 	& \multicolumn{2}{c|}{Complexity} \\
\cline{2-6}
 &PSNRYUV & VMAF  &  MS-SSIM & EncT &  DecT \\ 
\hline
\cellcolor{intra} MIP$^{\dagger}$	& 0.31\% & 0.57\% & 0.46\% & 96.0\% & 100.0\% \\ \hline
\cellcolor{intra} MRL$^{\dagger}$	&0.12\%&	0.16\%&	0.13\%&	100.0\%&	100.0\%\\ \hline
\cellcolor{intra} Lmchroma$^{\dagger}$	&5.07\%&	1.47\%&	1.31\%&	99.0\%&	100.0\%\\ \hline
\cellcolor{intra} ISP$^{\dagger}$	&0.36\%&	0.32\%&	0.33\%&	96.0\%&	100.0\%\\ \hline \hline
\cellcolor{tran} MTS$^{\oplus}$	&0.81\%&	1.07\%&	1.03\%&	93.7\%&	99.2\%\\ \hline
\cellcolor{tran} SBT$^{\oplus}$	&0.32\%&	0.52\%&	0.27\%&	95.0\%&	100.0\%\\ \hline
\cellcolor{tran} LFNST$^{\oplus}$ &	0.57\%&	0.97\%&	0.73\%&	96.1\%&	99.8\%\\ \hline
\cellcolor{tran} JCCR$^{\oplus}$	& 0.27\%&	0.38\%&	0.32\%&	99.0\%&	100.0\%\\ \hline
\cellcolor{tran} DQ$^{\oplus}$ &	2.15\%&	1.74\%&	2.20\%&	99.1\%&	100.0\%\\ \hline \hline
\cellcolor{inloop} DBF$^\otimes$	&0.44\%&	0.71\%&	0.20\%&	100.8\%&	85.0\%\\ \hline
\cellcolor{inloop} SAO$^\otimes$	& 0.08\%&	0.15\%&	0.02\%&	99.9\%&	98.1\%\\ \hline
\cellcolor{inloop} ALF+CCALF$^\otimes$	&7.64\%&	6.09\%&	0.66\%&	95.1\%&	90.0\%\\ \hline
\cellcolor{inloop} LMCS$^\otimes$&	0.15\%&	-1.01\%&	0.85\%&	97.6\%&	100.0\%\\ \hline \hline
\cellcolor{inter1}  TMVP$^\star$&	1.29\%&	1.56\%&	1.43\%&	99.4\%&	100.0\%\\ \hline
\cellcolor{inter1}  AMVR$^\star$	&1.29\%&	0.87\%&	1.16\%&	82.6\%&	100.0\%\\ \hline
\cellcolor{inter1}  MMVD$^\star$	& 0.28\%&	0.19\%&	0.23\%&	90.3\%&	100.0\%\\ \hline
\cellcolor{inter1}  SMVD$^{\star}$	&0.22\%&	0.19\%&	0.17\%&	96.2\%&	100.0\%\\ \hline  \hline 
\cellcolor{inter4} Affine$^\odot$ &	2.31\%&	2.32\%&	2.61\%&	80.8\%&	97.0\%\\ \hline
\cellcolor{inter4}  SBTMVP$^\odot$	&0.34\%&	0.53\%&	0.41\%&	101.1\%&	99.0\%\\ \hline \hline
\cellcolor{inter2}  BCW$^{\bullet}$	&0.25\%&	0.17\%&	0.15\%&	93.7\%&	98.0\%\\ \hline
\cellcolor{inter2} GPM$^{\bullet}$	&0.66\%&	0.73\%&	0.63\%&	95.1\%&	100.0\%\\ \hline
\cellcolor{inter2} CIIP$^{\bullet}$	&0.08\%&	0.17\%&	0.16\%&	96.7\%&	100.0\%\\ \hline \hline
\cellcolor{inter3} DMVR$^{\diamond}$ 	&0.86\%&	0.93\%&	1.00\%&	99.9\%&	95.8\%\\ \hline
\cellcolor{inter3} BDOF$^{\diamond}$ 	&0.64\%&	1.29\%&	0.74\%&	96.7\%&	98.0\%\\ \hline
\cellcolor{inter3} PROF$^{\diamond}$ 	&0.33\%&	0.41\%&	0.31\%&	98.3\%&	99.0\%\\
\hline
\end{tabular}
\end{adjustbox}
$^{\dagger}$~Intra tools, $^{\oplus}$~transform tools, $^\otimes$~in-loop filter tools, $^\star$~MV coding, $^\odot$~Subblock Motion Comp, $^{\bullet}$~motion compensation, $^{\diamond}$~prediction enhancement. \\
\end{table}
\begin{figure*}
    \centering
         \includegraphics[width=1\textwidth]{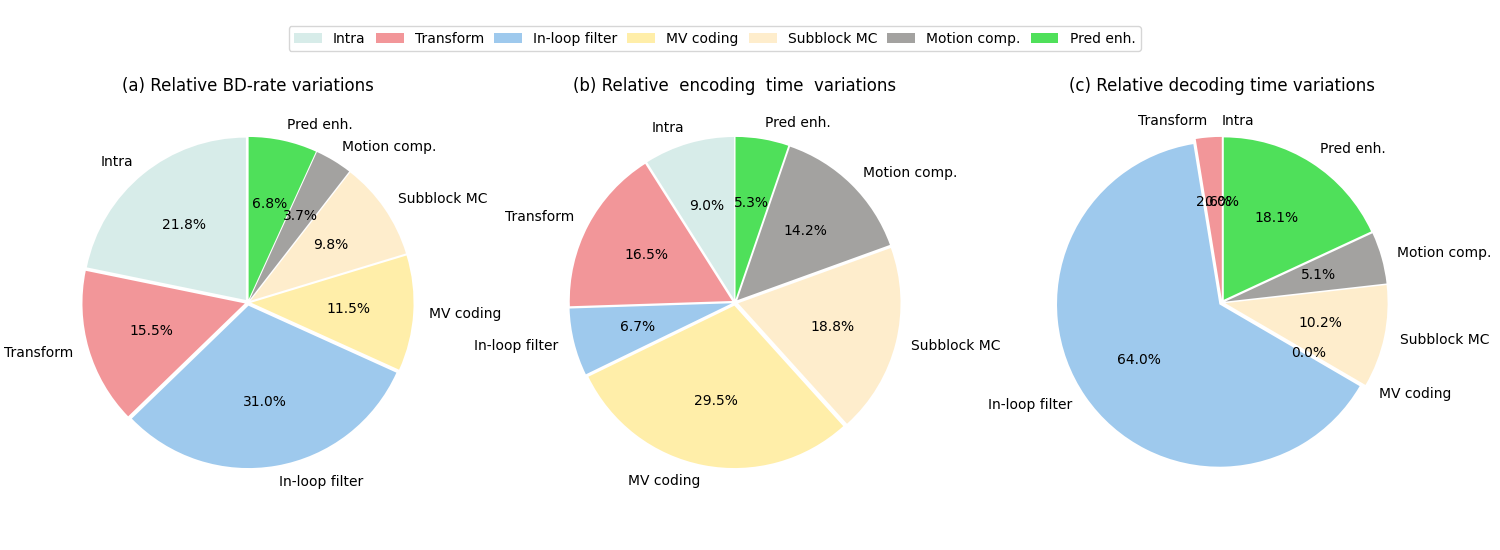}
    \caption{Relative contributions of the main \gls{vvc} tools' categories in \gls{ra} coding configuration.
    }
    \label{fig:my_label}
\end{figure*}

\addcomment{In} \Figure{\ref{fig:my_label}}, \addcomment{the tools listed in~\Table{\ref{table_example}} are grouped per category, and relative contribution of each category to the total PSNRYUV \gls{bd-br} coding gain brought by all the tools is depicted}. \addcomment{Similarly}, the \addcomment{relative} runtime ratios are depicted in \Figure{\ref{fig:my_label}}b and \Figure{\ref{fig:my_label}}c for encoder and decoder, respectively. All tools' categories turn out to have an important contribution to the overall coding gain. Most computing demanding parts in decoder are loop filtering and inter coding (motion compensation, sub-block MC and prediction enhancement). Intra and transform categories have negligible impacts on the \gls{vtm} decoding time. At \gls{vtm} encoding side, the most demanding part is the inter part, which represents 2/3rd of the encoding time increase. Another substantial part of the encoding runtime is consumed by the partitioning, which is not assessed and reported in this paper. The decoding runtime impact of these tools remains limited in this particular RA coding configuration~\cite{IBC2020}.


\section{Real-Time Implementations}
\label{sec:RTcodec}

As of now, \gls{vvc} benefits from both industrial and open-source implementations, contributing to the emergence of an end-to-end value chain. For example, manufacturers, universities and research-institutes have announced the availability of fast \gls{vvc} implementations~\cite{VVCDeployment}. Among existing solutions, the INSA Rennes real-time OpenVVC decoder, the Fraunhofer Heinrich Hertz Institute VVdeC decoder~\cite{9191199}, VVenC encoder~\cite{9287093} and the ATEME TitanLive encoding platform have been developed during the last months. 

\subsection{Real-Time \gls{vvc} Decoding with OpenVVC}
\label{sec:openVVC}
OpenVVC is the world-first real time software \gls{vvc} decoder developed from scratch in C programming language. The OpenVVC project is intended to provide consumers with an open source library enabling UHD real time \gls{vvc} decoding capability\footnote{\url{https://github.com/OpenVVC/OpenVVC}}. The \gls{vvc} main profile tools are supported by the OpenVVC decoder and the most complex operations are optimized in \gls{simd} for both Intel x86 and ARM Neon platforms. The decoder is parallel-friendly supporting high level parallelism such as parallel decoding of slices, tiles,  wavefront and frames (frame-based parallelism). These latter leverage multi-core platforms to further speedup the decoding process and reduce the decoded frame latency.    Finally, the decoder is compatible with the well known video players such as FFplay, GPAC and VLC.

\subsection{Real-Time \gls{vvc} Decoding with VVdeC}

It is important to consider that all the improvements added to \gls{vvc} in order to achieve better coding performance come with a cost in terms of computational complexity increasing,
which has been determined to be around 10x in the encoder and 2x in the decoder compared to \gls{hevc}. The Fraunhofer Heinrich Hertz Institute has been working on Versatile Video deCoder ~\cite{VVdeC} (VVdeC) since October 2020, aiming to provide a publicly available optimized \gls{vvc} software decoder\footnote{\url{https://github.com/fraunhoferhhi/vvdec}}. 

This implementation supports multicore architectures and it is optimized \addcomment{for x86 platforms}. This decoder takes advantage of functional (multi-threading) and data parallelization (\gls{simd} instructions) to achieve the maximum performance. Compared to \gls{vtm} decoder, this optimized implementation has reached 50\% to 90\% improvement in terms of decoding time reduction over a x86 platform ~\cite{CompDec}.

\subsection{Real-Time \gls{vvc} Encoding with VVenC}
VVenC~\cite{9287093} is an open source fast implementation \addcomment {of a} \gls{vvc} encoder\footnote{\url{https://github.com/fraunhoferhhi/vvenc}} developed by the Fraunhofer Heinrich Hertz Institute. VVenC is developed in C++ programming language and includes low level optimizations through \gls{simd} instructions targeting Intel x86 platform. The encoder is parallel-friendly enabling to process a set of pictures in parallel on multi-core platforms. Five presets are defined by the encoder offering a wide range trade-off between coding efficiency (quality) and speed (complexity). Perceptual optimization is also integrated to improve subjective video quality, based on the XPSNR visual model~\cite{9054089}. Finally, frame-level single-pass and two-pass rate control with variable bit-rate (VBR) encoding are supported by the encoder.

\subsection{Real-Time \gls{vvc} Encoding and Packaging with TitanLive}

The ATEME TitanLive solution provides software-based implementation of a wide variety of standards for audio/video coding, packaging and transport. This solution is currently used worldwide for broadcast and \gls{ott} head-end deployments. In order to support \gls{vvc}, a number of components were upgraded. 

As further described in~\cite{VVC8K}, \gls{vvc} and \gls{hevc} present some structural similarities making an upgrade from \gls{hevc} to \gls{vvc} feasible in a cost effective manner. In order to do so, the \gls{vvc} syntax has been implemented with support for the tools already implemented in \gls{hevc}, disabling the other ones in the \gls{aps}. Then, the \gls{hevc} tools have been upgraded to comply with \gls{vvc} specification and some tools offering a good complexity-vs-gains trade-offs were implemented. Relying on the same core coding engine enabled us to leverage the existing optimized function (assembly, intrinsic) to achieve \gls{vvc} real-time encoding with interesting gains over \gls{hevc}, from 10\% to 15\% depending on the video content.

The packager has been upgraded as well to support \gls{vvc} encapsulation into MPEG2-TS and ISOBMFF. Since the final draft international standard (FDIS) has not been issued yet for MPEG2-TS and ISOBMFF binds, some draft amendment versions (DAM) were implemented~\cite{mpegts_vvc,isobmff_vvc}, assuming that these versions are close to the FDIS ones.

\begin{figure*}[t]
\begin{subfigure}[t]{.9\linewidth}
\centering
\includegraphics[width=\linewidth]{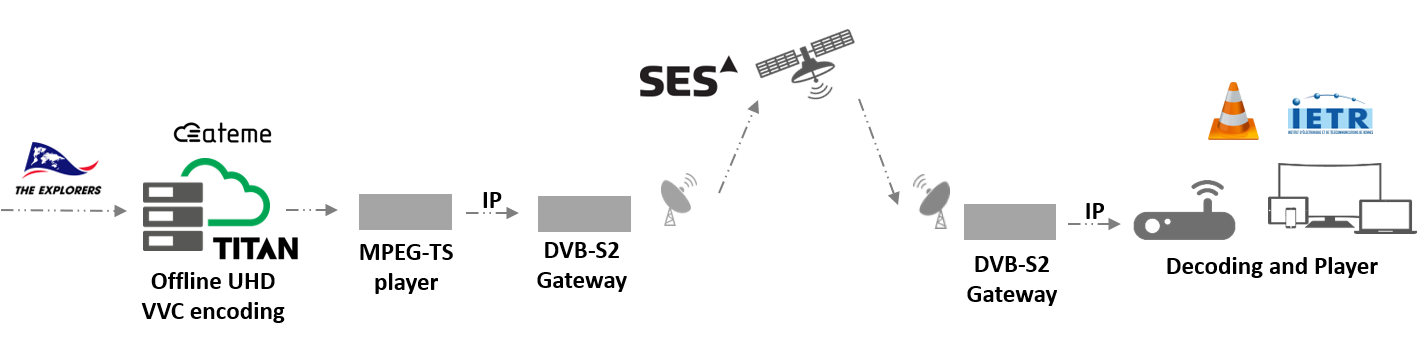}

\caption{End-to-end over the air video broadcasting with \gls{vvc}.} \label{fig:end2end_ot1}
\vspace{10pt}
\end{subfigure}
  \begin{subfigure}[t]{.94\textwidth}
  \centering
    \includegraphics[width=.94\textwidth]{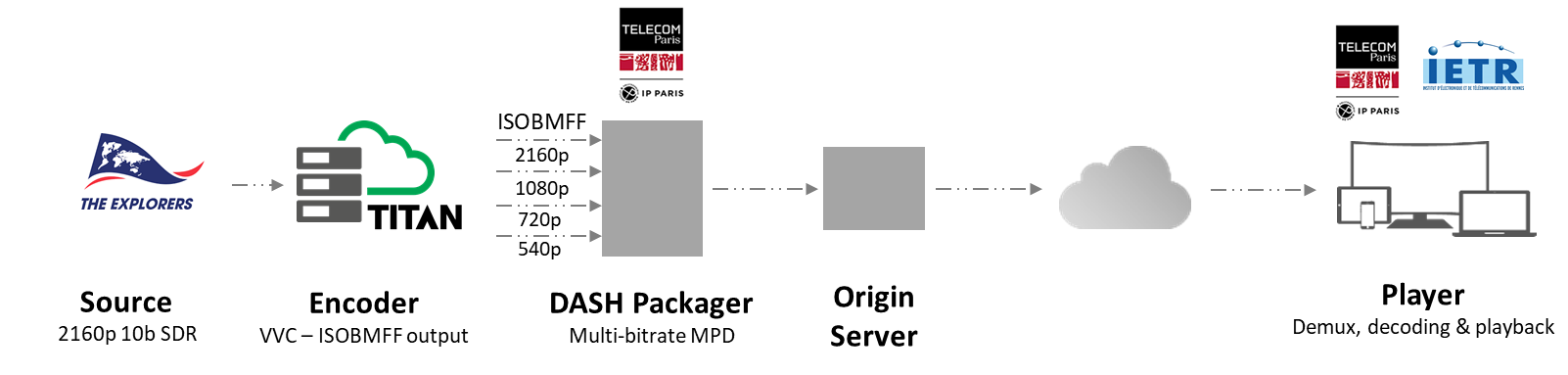}
\caption{End-to-end \gls{ott} video delivery with \gls{vvc}.}\label{fig:end2end_ot2}
  \end{subfigure}\hfill
  \begin{subfigure}[t]{.9\textwidth}
  \centering
    \includegraphics[width=.94\textwidth]{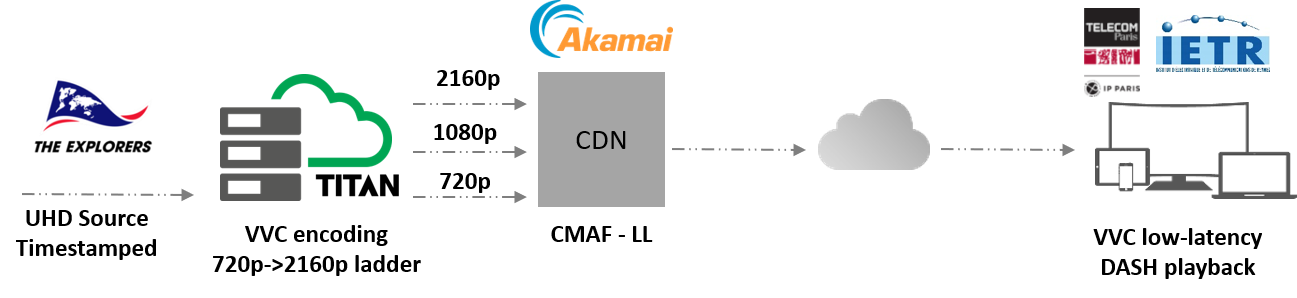}
    \caption{Live and Low-Latency 4K video \gls{ott} channel using \gls{vvc}.} \label{fig:end2end_ot3}
  \end{subfigure}
\caption{Illustration of different end-to-end demonstrations with the \gls{vvc} standard on different transmission mediums.}
\label{fig:blocks}
\vspace{-0.2in}
\end{figure*}
\section{First Commercial Trials}
\label{sec:trials}

Using the previously described tools some  world-first \gls{vvc} in field trials ~\cite{worldfirst_ota,worldfirst_ott} have been implemented as described in the following subsections. 

\subsection{World-First OTA Broadcasting with \gls{vvc}}

\subsubsection{Overview}

The trial depicted in Figure~\ref{fig:end2end_ot1} took place in June 2020 and is the result of a collaboration between the following entities:
\begin{itemize}
    \item ATEME provided the encoding and packaging units.
    \item SES provided the satellite transponder used for the experiment as well as the gateways needed at transmission and reception sides.
    \item VideoLabs provided the media player (VLC) including demuxing and playback.
    \item IETR provided the \gls{vvc} real-time decoding library used by VLC player.
\end{itemize}
As illustrated, the UHD source provided by The Explorers is encoded with \gls{vvc}, and encapsulated in MPEG-TS using ATEME video processing platform. \addcomment{The provided video bitstreams were received by SES and sent to the modulator gateways feeding the ASTRA 2E transponder (Europe coverage).} The signal is demodulated and forwarded on IP to a VLC player that displays the video thanks to the real-time OpenVVC decoder developed by IETR.

\subsubsection{\gls{vvc} Encoding and Encapsulation}

The ATEME encoding engine used in this experiment followed the \gls{vvc} draft specification~\cite{vvc_6}  and produced a bitstream decodable by the \gls{vtm} software (tag version 6.1). The 2160p-10b-SDR video input was encoded offline using \gls{ra} \gls{gop} structure with a 1sec RAP period and a 20Mbps constant bitrate (CBR) and deltaQP is enabled.

The produced elementary stream was encapsulated in MPEG-TS following draft specification incorporating \gls{vvc} amendments~\cite{isobmff_6}. Hence, a stream embedding stream\_type 0x32 with video descriptor 57 was generated and ready to be delivered over existing broadcast infrastructure.

\subsubsection{Satellite Transmission}

The MPEG-TS provided by ATEME was rate-adapted and played out on Transponder 2.014 on SES prime UK position 28.2 East. The transponder used for this transmission is a transponder that has been used previously for SES 8K test transmissions and therefore the parameters and link budget were slightly unusual for transmissions at this position (Freq: 11.973, Pol: Vertical, SR 31 MS/s, \gls{dvb}-S2, 8PSK 9/10). The uplink was done in Betzdorf, Luxembourg which is the location of SES headquarters.

Reception was done using a \gls{dvb}-S2 to IP Gateway (in our instance the Kathrein EXIP 4124 which is a SAT$>$IP server). The Gateway was statically tuned to the corresponding transponder and setup to forward the relevant PIDs of the received TS encapsulated in RTP multicast (RFC 2250) to the local network. On that network a powerful PC was used to decode the 4K-UHD stream in real-time and displayed on several TVs. The PC framerate was set to the framerate of the source content.

\subsubsection{Player and Decoder}

The OpenVVC described in Section~\ref{sec:openVVC} was used as decoding library for this trial. The VLC player wraps both OpenVVC and libavformat demuxer for this trial. Libavformat demuxer is modified to handle MPEG-TS streams carrying \gls{vvc} and the extracted NALUs are processed by OpenVVC. The synchronization and picture presentation is managed by VLC player.


\subsection{World-First \gls{ott} Delivery with \gls{vvc}}

\subsubsection{Overview}
The trial depicted in Figure~\ref{fig:end2end_ot2} took place in June 2020 and is the result of a collaboration between the following entities:
\begin{itemize}
    \item ATEME provided the encoding unit.
    \item Telecom Paris provided the DASH packager (MP4Box) and the player (MP4Client) from GPAC.
    \item IETR provided the \gls{vvc} real-time decoding library used by MP4Client.
\end{itemize}
As illustrated, the UHD source provided by The Explorers is encoded with \gls{vvc}, and formatted into ISOBMFF mp4 files using ATEME video processing platform. The mp4 files are encapsulated into DASH with multiple representations using Telecom Paris MP4Box software. The generated DASH is pushed on an origin server publicly accessible on the internet. The MP4Client demultiplexes and plays the content thanks to the real-time OpenVVC decoder developed by IETR.

\subsubsection{\gls{vvc} Encoding}

The ATEME encoding engine used in this experiment followed the \gls{vvc} draft specification~\cite{vvc_6} and produced a bitstream decodable by the \gls{vtm} software (tag version 6.1). The video input was encoded offline using \gls{ra} closed-\gls{gop} structure with a 1sec RAP period, producing the following bitrate ladder:
\begin{itemize}
    \item 540p @ 1.6 Mbps
    \item 720p @ 3.4 Mbps
    \item 1080p @ 5.8 Mbps
    \item 2160p @ 16.8 Mbps
\end{itemize}

\subsubsection{Client and Player}
Support for \gls{vvc} transport was added to GPAC~\cite{le_feuvre_gpac_2007} as follows:
\begin{itemize}
    \item ISOBMFF demultiplexing for 'vvc1' and 'vvi1' sample description entries has been added
    \item ISOBMFF multiplexing for 'vvc1' and 'vvi1' sample description entries has been added
    \item Inspection of files containing \gls{vvc} tracks has been added (partial support, no bitstream parsing of \gls{vvc} has been added yet).
\end{itemize}

The tools used in this demo were based on GPAC 1.0. In this version, the DASH segmenter is independent from the media packaging format and did not require any modification. It will however require further update once the "codecs" MIME parameter for \gls{vvc} have been defined; for the purpose of the demo, the "codecs" MIME parameter for \gls{vvc} is set to "vvc1" only. Since the demo relies on GPAC 1.0, the packager can output to both MPEG-DASH and HLS formats.

Similarly, the DASH access engine in GPAC is independent from the media packaging format and did not require any modification. The OpenVVC decoder is integrated through a patched version of libavcodec. The playback chain has been tested under Windows, Linux and Mac OSX platforms. Since the experiment, the \gls{vvc} support has been merged in the GPAC's main code repository in the master branch\footnote{\url{https://github.com/gpac/gpac/tree/master}}.



\subsection{World-First 4K Live \gls{ott}Channel with VVC}

The trial depicted in Figure~\ref{fig:end2end_ot3} took place in September 2020 and is the result of a collaboration between the following entities:
\begin{itemize}
    \item ATEME provided the encoding unit.
    \item Telecom Paris provided the DASH packager (MP4Box) and the player (MP4Client) from GPAC.
    \item IETR provided the \gls{vvc} real-time decoding library used by MP4Client.
    \item Akamai provided \gls{cdn} infrastructure supporting HTTP chunk transfer encoding to enable low-latency.
\end{itemize}

In this experiment, an end-to-end live 4K TV channel was demonstrated during a period of 1 month. The input video, provided by The Explorers was timestamped with UTC time and live-encoded using the TitanLive platform. The \gls{vvc} encoding was carried out with low-latency CMAF packaging, issuing 100ms chunks within a 2000ms segments, pushed to the Akamai \gls{cdn} thanks to HTTP chunk transfer encoding. The GPAC player described in the previous section was used and highlighted the low-latency delivery when compared to UTC time at the receiver side (typically measuring a 2s glass-to-glass).

\section{Conclusion}
\label{sec:conc}
In this paper, we have addressed several important aspects of the latest video coding standard \gls{vvc} from market use-cases, coding tools description, per-tool coding efficiency and complexity assessments, to the description of real time implementations of \gls{vvc} codecs used in early commercial trials. \addcomment{Real-time implementations of} \gls{vvc} \addcomment{codecs as well as its adoption by application standards are essential} to ensure a wide adoption and a successful deployment of the \gls{vvc} standard. The current status of the developed real time \gls{vvc} codecs and the demonstrated end-to-end \gls{vvc} transmissions over broadcast and \gls{ott} communication mediums clearly show that the \gls{vvc} technology is mature enough and ready for real deployment on consumer electronic products. Our prediction is that \gls{vvc} will be integrated in most of the consumer electronics devices in a near future.





\bibliographystyle{IEEEtran}
\bibliography{IEEEexample, ref}

%





\end{document}